\documentclass[useAMS,usenatbib,twocolumn]{mn2e}
\usepackage[dvips]{graphicx}
\usepackage{amsmath}
\usepackage{amsfonts}
\usepackage{amssymb}

\title[Investigations of transit duration expressions]{Investigations of approximate expressions for the transit duration}
\author[D. M. Kipping]{David M. Kipping$^{1,2}$\thanks{E-mail:
d.kipping@ucl.ac.uk}\footnotemark[1] \\
$^{1}$Harvard-Smithsonian Center for Astrophysics, 60, Garden St., \\
       Cambridge, MA 02138, USA \\
$^{2}$Department of Physics and Astronomy, University College London, \\
       Gower Street, London WC1E 6BT, UK}
\begin{document}

\date{Accepted . Received ; in original form }

\pagerange{\pageref{firstpage}--\pageref{lastpage}} \pubyear{2010}
\maketitle

\label{firstpage}

\begin{abstract}

In this work, we investigate the accuracy of various approximate expressions for the transit duration of a detached binary against the exact solution, found through solving a quartic equation.  Additionally, a new concise approximation is derived, which offers more accurate results than those currently in the literature. Numerical simulations are performed to test the accuracy of the various expressions.  We find that our proposed expression show yields a $>200$\% improvement in accuracy relative to the most previously employed expression.

We derive a new set of equations for retrieving the lightcurve parameters and consider the effect of falsely using circular expressions for eccentric orbits, with particularly important consequences for transit surveys. The new expression also allows us to propose a new lightcurve fitting parameter set, which minimizes the mutual correlations and thus improves computational efficiency.  The equation is also readily differentiated to provide analytic expressions for the transit duration variation (TDV) due to secular variations in the system parameters, for example due to apsidal precession induced by perturbing planets.

\end{abstract}


\section{Introduction}

Transiting exoplanets and eclipsing binaries produce familiar U- and V-shaped lightcurves with several defining quantities, such as the mid-eclipse time, eclipse depth and duration. Out of these, the transit duration is undoubtedly the most difficult observeable to express in terms of the physical parameters of the system. Kipping (2008) showed that the duration is found by solving a quartic equation, to which exists a well-known solution. In general, two roots correspond to the primary eclipse and two to the secondary but this correspondence determination has an intricate dependency on the input parameters for which there currently exists no proposed rules. As a consequence, there currently exists no single exact expression for the transit duration.

In many applications, the process of discarding unwanted roots may be performed by a computer, but naturally this can only be accomplished for case-by-case examples. The benefits of a concise, accurate and general expression for the transit duration, as we will refer to it from now on, are manifold.  The solution provides lower computation times, deeper insight into the functional dependence of the duration and a decorrelated parameter set for fitting eclipse lightcurves (see \S6).  Such a solution may also be readily differentiated to investigate the effects of secular and periodic changes in the system (see \S7). With the changes in transit duration recently being proposed as a method of detecting additional exoplanets in the system (Miralda-Escud\'{e} 2002; Heyl \& Gladman 2007) and companion exomoons (Kipping 2009a,b), there is a strong motivation to ensure an accurate, elegant equation is available.

In this work, we will first propose a new approximate expression for the transit duration in \S2. In \S3, we will derive two new approximate expressions for the transit duration and discuss others found in the literature; exploring their respective physical assumptions and derivations. In \S4, we present the results of numerical tests of the various formulae.  We find that one of our new proposed expressions offers the greatest accuracy out of the candidates.  In \S5 and \S6, we utilize the new equation to derive mappings between the observeable transit durations and the physical model parameters.  These mappings may be used to obtain a decorrelated transit lightcurve fitting parameter set, which enhances the compuational efficiency of Monte Carlo based methods. Finally, in \S7, we differentiate the favoured solution to predict how the transit duration will change due to apsidal precession, nodal precession, in-fall and eccentricity variation.

\section{The Transit Duration Equation}
\subsection{Exact solution}

Before we begin our investigation, let us first clearly define what we mean by the transit duration.  There exists several different definitions of what constitutes as the transit duration in the exoplanet literature.  In figure 1, we present a visual comparison of these definitions.  There exists at least 7 different duration definitions: $t_T$, $t_F$, $T$, $W$, $\tau$, $t_{12}$ and $t_{34}$.  $t_T$ and $t_F$ are the definitions used by \citet{sea03} and represent the $1^{\mathrm{st}}$ to $4^{\mathrm{th}}$ and $2^{\mathrm{nd}}$ to $3^{\mathrm{rd}}$ contact durations respectively.  These can be understood to be the total transit duration and the flat-bottomed transit duration (sometimes confusingly referred to as the full duration).  $T$ is the time for a planet to move from its sky-projected centre overlapping with the stellar limb to exitting under the same condition.  $W$ is a parameter we define in this work as the average of $t_T$ and $t_F$ which we call the transit width.  We note that $T > W$ since when the planet's centre crosses the limb of star, less than one half of the planet's projected surface blocks the light from the star.  However, many sources in literature make the approximation $T \simeq W$, which is only true for a trapezoid approximated lightcurve (a detailed discussion on this is given in \S6.3).  Finally, $t_{12}$ and $t_{34}$ are the ingress and egress durations respectively which are often approximated to be equal to one single parameter, $\tau$.

\begin{figure}
\begin{center}
\includegraphics[width=8.4 cm]{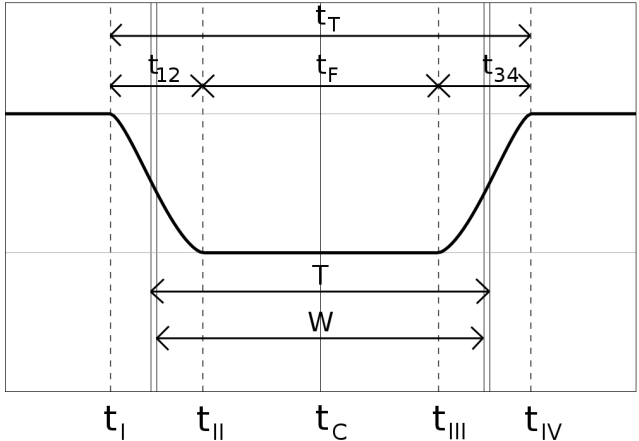}
\caption{\emph{Various definitions of the transit duration are marked on a model transit lightcurve.}} \label{fig:fig1}
\end{center}
\end{figure}

Throughout this paper, we assume that the planet is a black sphere emitting no flux crossing the disc of a perfectly spherical star.  The consequences for oblate planets is discussed by \citet{sea02} and for hot planets with significant nightside fluxes in \citet{kip10}.  We will employ the definition of $T$ for the transit duration.  Once the equation for $T$ is known, it is trivial to transmute it to give any of the other definitions provided in figure 1.

An exact solution for $T$, in terms of the true anomaly $f$, is given by integrating d$f$/d$t$ between $f_b$ and $f_a$ (where we use $f$ to describe true anomaly throughout this paper). Details of the derivation can be found in K08, but to summarize we have:

\begin{align}
&T(f_b,f_a) = \Big(\frac{P}{2 \pi} \frac{1}{\sqrt{1-e^2}}\Big) \cdot \Big[ D(f_b) - D(f_a) \Big] \\
&D(f) = 2 \sqrt{1-e^2} \tan^{-1} \Big[\sqrt{\frac{1-e}{1+e}} \tan{\frac{f}{2}}\Big] - \frac{e (1-e^2) \sin f}{1+e \cos f} \\
&D(f_b) - D(f_a) = 2 \sqrt{1-e^2} \tan^{-1}\Bigg[\frac{\sqrt{1-e^2} \sin f_H}{\cos f_H+e \cos f_M}\Bigg] \nonumber \\
\qquad& -\frac{2 e (1-e^2) \sin f_H (e \cos f_H + \cos f_M)}{(1-e^2) \sin^2f_M + (e \cos f_H+\cos f_M)^2}
\end{align}

Where $P$ is the planetary orbital period, $e$ is the orbital eccentricity, $D(f)$ is the `duration function', $f_M = (f_b + f_a)/2$ and $f_H = (f_b - f_a)/2$. It is clear that two principal terms define $D$ and hence we dub the expressions used here as the `two-term' transit duration equation. At this point, the outstanding problem is solving for $f_a$ and $f_b$, which are the true anomalies at the contact points.

The K08 solution is derived using Cartesian coordinates, but we consider here a simpler formulation in terms of $\cos(f)$. In order to solve, we must first consider the geometry of the system.  As with K08, an ellipse is rotated for argument of pericentre and then orbital inclination.  We choose to define our coordinate system with the star at the origin with the $+Z$ axis pointing at the observer.  We also choose to align the $+X$ axis towards the ascending node of the planetary orbit, which ensures that the longitude of the ascending node satisfies $\Omega = 0$ and the longitude of pericentre (of the planetary orbit) is equal to the argument of the pericentre ($\varpi = \omega$).  In such a coordinate system, we may write the sky-projected planet-star separation (in units of stellar radii):

\begin{equation}
S(f) = a_R \frac{1-e^2}{1+e \cos f} \sqrt{1- \sin^2i\sin^2(\omega+f)}
\end{equation}

Where $a_R$ is the semi-major axis in units of stellar radii ($a/R_*$) and $i$ is the orbital inclination. The two contact points occur when $S$ is equal to unity.  For a body which undergoes both primary and secondary transit, there must be at least four solutions, which already is an indication of a quartic.  If we let $c = \cos f$ and rearrange (4) in terms of purely terms in $c$, we obtain the following quartic equation:

\begin{align}
0 =& Q_0 + Q_1 c + Q_2 c^2 + Q_3 c^3 + Q_4 c^4 \\ 
Q_0 =& (\csc^2i+\Lambda^2 (\cos^2\omega-\csc^2i))^2 \\
Q_1 =& 4e \csc^2 i (\csc^2+\Lambda^2(\cos^2\omega-\cos^2i)) \\
Q_2 =& \Lambda^2 \csc^2i (e^2+(2\Lambda^2+e^2-2) \cos2\omega) \nonumber \\
\qquad&+2e^2 \csc^4i(3-\Lambda^2)-2\Lambda^4 \cos^2\omega \\
Q_3 =& 4 e \csc^2i (e^2 \csc^2i-\Lambda^2 \cos2\omega) \\
Q_4 =& (\Lambda^2 \cos2\omega-e^2 \csc^2i)^2 + \Lambda^4 \sin^22\omega \\
\Lambda =& a_R (1-e^2) 
\end{align}

Equation (5) is a quartic equation not satisfying any of the special case quartics which are most easily solved (e.g. a biquadratic).  Since we have four roots for $c=\cos f$, there are eight roots in total for $f$, of which only four are physical. Therefore it is preferable to always work with $\cos f$ since $D(f)$ may be easily expressed in terms of $c$ as well. Although the solutions of a quartic equation are well known, two of the roots correspond to the primary transit and two to the secondary. The correspondence of which roots relate to which contact points varies with an intricate dependency on the input parameters. Unfortunately, no known rules or relations currently exist for this correspondence and we were unable to find a system either. Consequently, there currently exists no single exact equation for the transit duration.

We note that the problem of finding the duration of an eclipse is not a new one. An equivalent problem is considered in \citet{kop59} for the time taken for a body to move between primary and secondary transit. \citet{kop59} showed that an closed-form expression is possible by assuming $i=90^{\circ}$. However, this assumption would be too erroneous for the purposes of finding the duration of a transit event.

\subsection{Approximation Solutions}

In order to avoid the quartic equation, we must make an approximation.  A useful approximation we can make is that $\varrho(f) \simeq \varrho_c = \varrho(f=f_c)$, i.e. the planet-star separation is approximately a constant value given by the planet-star separation at mid-transit.  Defining the transit impact parameter as $b = a_R \varrho_c \cos i$, it may be shown that the difference between $f_b$ and $f_a$ is given by:

\begin{equation}
\sin f_H = \sin\Big(\frac{f_b - f_a}{2}\Big) \simeq \frac{\sqrt{1-b^2}}{a_R \varrho_c \sin i}
\end{equation}

In addition to (12), we require $(f_a + f_b)/2$. A good approximation would appear to be that $(f_b + f_a)/2 \simeq f_c$, which is the true anomaly at mid-transit.  $f_c$ is defined as the point where $S$ is a minimum.  Differentiating $S$ with respect to $f$ and solving for $f$ leads to a quartic expression again and thus an exact concise solution remains elusive but a good approximation is given by:

\begin{equation}
f_M \simeq f_c \simeq \frac{\pi}{2} - \omega
\end{equation}

\subsection{Two-Term Expression}

By combining equations (1) \& (3) with (12) \& (13), we are able to obtain a final expression for the duration, which we dub $T_{2}$ (`two-term').  Testing $T_2$ for the exact solutions for $f_M$ and $f_H$ provided precisely the correct transit duration for all $e$, as expected.  However, we found that using approximate entries for these terms severely limited the precision of the derived equation for large $e$ (the results of numerical tests will be shown later in \S4).

The source of the problem comes from equation (3) which consists of taking the difference between two terms.  Both terms are of comparable magnitude for large $e$ and thus we are obtaining a small term by taking the difference between two large terms.  These kinds of expressions are very sensitive to slight errors.  In our case, the error is from using approximate entries for $f_M$ and $f_H$.  In this next section, we will consider possible `one-term expressions' which avoid the problem of taking the difference of two comparable-magnitude terms.

\subsection{One-Term Expression}

There are numerous possible methods for finding `one-term' transit duration expressions.  Starting from equation (1), we could consider using the same assumption which we used to derive the approximate true anomalistic duration, $\Delta f$; i.e. the planet-star separation does not change during the transit event.  This would yield:

\begin{align}
T_1 &= \frac{P}{2 \pi} \frac{\varrho_c^2}{\sqrt{1-e^2}} \Delta f \\
T_1 &= \frac{P}{\pi} \frac{\varrho_c^2}{\sqrt{1-e^2}} \arcsin\Bigg(\frac{\sqrt{1-a_R^2 \varrho_c^2 \cos^2i}}{a_R \varrho_c \sin i}\Bigg)
\end{align}

Where we have used equation (12) for $\Delta f$.  Another derivation would be to assume the planet takes a tangential orbital velocity and constant orbital separation from the planet, sweeping out an arc of length $r_c \Delta f$.  It is trivial to show that this argument will lead to precisely the same expression for $T_1$.

\section{Current Expressions for the Transit Duration}
\subsection{\citet{sea03} Equation}

For a circular orbit, the task of finding the transit duration is greatly simplified due to the inherent symmetry of the problem and an exact, concise solution is possible, as first presented by \citet{sea03} (SMO03).

\begin{equation}
T_{\mathrm{SMO03}} = \frac{P}{\pi} \arcsin\Bigg(\frac{\sqrt{1-a_R^2 \cos^2i}}{a_R \sin i}\Bigg)
\end{equation}

The physical origin of this expression can be seen as simply multiplying the reciprocal of the planet's tangential orbital velocity (which is a constant for circular orbits), by the distance covered over the swept-out arc, $a \Delta f$.

\begin{align}
T_{\mathrm{SMO03}} &= v^{-1} \cdot \Delta d \nonumber \\
T_{\mathrm{SMO03}} &= \Big(\frac{P}{2 \pi a}\Big) \cdot (a \Delta f)
\end{align}

Where we expand the first term as the orbital period divided by the orbital circumference, and the second term as the arc length.  It may be shown that:

\begin{equation}
\Delta f(e=0) = 2 \arcsin\Bigg(\frac{\sqrt{1-a_R^2 \cos^2i}}{a_R \sin i}\Bigg)
\end{equation}

It can be seen that our approximate expression for $\Delta f$, presented in equation (12) is equivalent to equation (18) for circular orbits.  It is also worth noting that both $T_1$ and $T_2$ can be shown to reduce down to $T_{\mathrm{SMO03}}$ for circular orbits.

\subsection{\citet{tin05} Equation}

\citet{tin05} (TS05) presented expressions for the duration of an eccentric transiting planet, which has been used by numerous authors since (e.g. Ford et al. 2008; Jord\'{a}n \& Bakos 2008).  It is also forms the basis of a lightcurve parameter fitting set proposed by Bakos et al. (2007). There are two critical assumptions made in the derivation of the TS05 formula.  The first of these is that:

\begin{itemize}
\item The planet-star separation, $r$, is constant during the planetary transit event and equals $r_c$
\end{itemize}

This is the same assumption made in the derivation of the $T_1$ equation.  Under this assumption, TS05 quote the following expression for $T$ (changing to consistent notation).

\begin{equation}
T_{\mathrm{TS05}} = \frac{r_c \Delta \phi}{v_c}
\end{equation}

Where TS05 define $r_c$ as the planet-star separation at the moment of mid-transit, $v_c$ as the planet's orbital velocity at the moment of mid-transit and $\Delta \phi$ as `the eccentric angle between the first and last contacts of transit'. In the standard notation, there is no such parameter defined strictly as the `eccentric angle' and thus we initially assumed that TS05 were referring to the eccentric anomaly.  However, substiuting the revevant terms for $r_c$ and $v_c$ gives:

\begin{equation}
T_{\mathrm{TS05}} = \frac{P}{2 \pi} \frac{\varrho_c^2}{\sqrt{1-e^2}} \Delta \phi
\end{equation}

By comparing (20) to equation (14), it is clear $\Delta \phi = \Delta f$ (also note equation (14) was derived under precisely the same assumptions as that assumed by TS05 at this stage of the derivation).  We therefore conclude that the term TS05 refer to as `eccentric angle' infact refers to true anomaly.  This is an important point to make because the derivation of the TS05 equation would otherwise be very difficult to understand by those working outside of the field.  Continueing the derivation from this point, the second assumption made by TS05 is:

\begin{itemize}
\item The planet-star separation is much greater than the stellar radius, $r \gg R_*$
\end{itemize}

Critically, this assumption was not made in the derivation of $T_1$ or $T_2$.  Using this assumption, TS05 propose that (replacing $\Delta\phi \rightarrow \Delta f$ to remain consistent with the notations used in this work and replacing $(1+p) \rightarrow 1$ to refer to $T$ rather than the duration from contact points 1 to 4):

\begin{align}
\Delta f_{\mathrm{TS05}} &= \arcsin \Bigg(2 \frac{\sqrt{1 - a_R^2 \varrho_c^2 \cos^2i}}{a_R \varrho_c}\Bigg) \\
\Delta f_{\mathrm{TS05}} &\simeq 2 \frac{\sqrt{1 - a_R^2 \varrho_c^2 \cos^2i}}{a_R \varrho_c}
\end{align}

Where TS05 use equation (22) rather than (21) in the final version of $T$.  Therefore, TS05 effectively make a small-angle approximation for $\Delta f$, which is a knock-on effect of assuming $r \gg R_*$.  We argue here that losing the $\arcsin$ function does not offer any great simplification of the transit duration equation but does lead to an unneccessary source of error in the resultant expression, in particular for close-in orbits, which is common for transits.  We also note that even equation (21) exhibits differences to equation (12).

Firstly, inside the $\arcsin$ function, the factor of $\csc i$ is missing which is present in both the derivation we presented in equation (12) and the derivation of SMO03 for circular orbits, equation (18).  The absence of this term can be understood as a result of the $r \gg R_*$ assumption. As $r \rightarrow \infty$, in order to maintain a transit event, we must have $i \rightarrow (\pi/2)$.

Secondly, the expression we presented for $\Delta f$ earlier in (12) has the factor of 2 present outside of the arcsin function, whereas TS05 have this factor inside the function.  Furthermore, the SMO03 derivation also predicts that the factor of 2 should be outside of the $\arcsin$ function and this expression is known to be an exact solution for circular orbits.  In a small angle approximation, $\arcsin2 x \simeq 2 \arcsin x$, but we point out that moving the factor of 2 to within the $\arcsin$ function seems to serve no purpose except to invite further error into the expression.  As a result of these differences, the TS05 expression for $T$ does not reduce down to the original SMO03 equation and is given by:

\begin{equation}
T_{\mathrm{TS05}} = \frac{P}{\pi} \frac{\varrho_c}{\sqrt{1-e^2}} \frac{\sqrt{1 - a_R^2 \varrho_c^2 \cos^2i}}{a_R}
\end{equation}

\subsection{\citet{win10} Equation}

\citet{win10} (W10) proposed an expression for $T$ based on modification to the SMO03 equation.  The first change was to modify the impact parameter from $a_R \cos i \rightarrow \varrho_c a_R \cos i$, i.e. to allow for the altered planet-star separation for eccentric orbits.  Secondly, the altered planetary velocity should also be incorporated.  W10 propose that a reasonable approximation for the transit duration is obtained by multiplying the SMO03 expressions by the following ratio:

\begin{equation}
\frac{\frac{\mathrm{d}X}{\mathrm{d}t}(f_c)[e=0]}{\frac{\mathrm{d}X}{\mathrm{d}t}(f_c)} = \frac{\varrho_c}{\sqrt{1-e^2}}
\end{equation}

Where $X$ is given in \citet{win10} and $f_c$ is the true anomaly at the centre of the transit.  This yields a new transit duration equation of:

\begin{equation}
T_{\mathrm{W10}} = \frac{P}{\pi} \frac{\varrho_c}{\sqrt{1-e^2}} \arcsin\Bigg(\frac{1}{a_R} \Big[\frac{1-a_R^2 \varrho^2 \cos^2 i}{\sin^2 i}\Big]^{1/2}\Bigg)
\end{equation}

Firstly, we note an obvious improvement of the W10 expression is that we recover the original SMO03 equation for $e=0$.  Secondly, comparison to the equation for $T_1$ reveals that the two expressions are very similar except for the position of an extra $\varrho_c$ term.  Indeed, the $T_1$ and $T_{\mathrm{W10}}$ expressions are equivalent in the small-angle approximation.

\section{Numerical Investigations}
\subsection{Example Systems}

Insights into the robustness and accuracy of the various expressions may be obtained through numerical tests of the various approximate expressions. We here compare the accuracy of four expressions: $T_{\mathrm{TS05}}$, $T_{\mathrm{W10}}$, $T_1$ and $T_2$. These expressions depend only on five parameters $P$, $a/R_*$, $b$, $e\sin\omega$ and $e\cos\omega$. One of the clearest ways of visually comparing the equations is to consider a typical transiting exoplanet example with system parameters for $a/R_*$ and $b$ and vary the eccentricity parameters. $P$ may be selected by simply assuming a star of Solar density. 

We create a 1000 by 1000 grid of $e\sin\omega$ and $e\cos\omega$ values from -1 to 1 in equal steps. Grid positions for hyperbolic orbits ($e > 1$) are excluded. We then calculate the transit duration through the exact solution of the quartic equation, $T_{\mathrm{K08}}$, plus all four approximate formulae.  We then calculate the fractional deviation of each equation from the true solution using:

\begin{equation}
\mathcal{D}_{\mathrm{candidate}} = \frac{T_{\mathrm{candidate}} - T_{\mathrm{K08}}}{T_{\mathrm{K08}}}
\end{equation}

We then plot the loci of points for which the deviation is less than 1\% (i.e. $\mathcal{D}_{\mathrm{candidate}} < 0.01$). In figure 2, we show four such plots for different choices of $a/R_*$ and $b$. The plot reveals several interesting features:

\begin{itemize}
\item $T_1$ consistently yields the largest loci.
\item $T_2$ is sometimes accurate and sometimes not, supporting the hypothesis that the approximation is not stable.
\item $T_{\mathrm{W10}}$ also yields consistently large loci.
\item $T_{\mathrm{TS05}}$ consistently yields the smallest loci.
\end{itemize}


\begin{figure*}
\begin{center}
\includegraphics[width=16.8 cm]{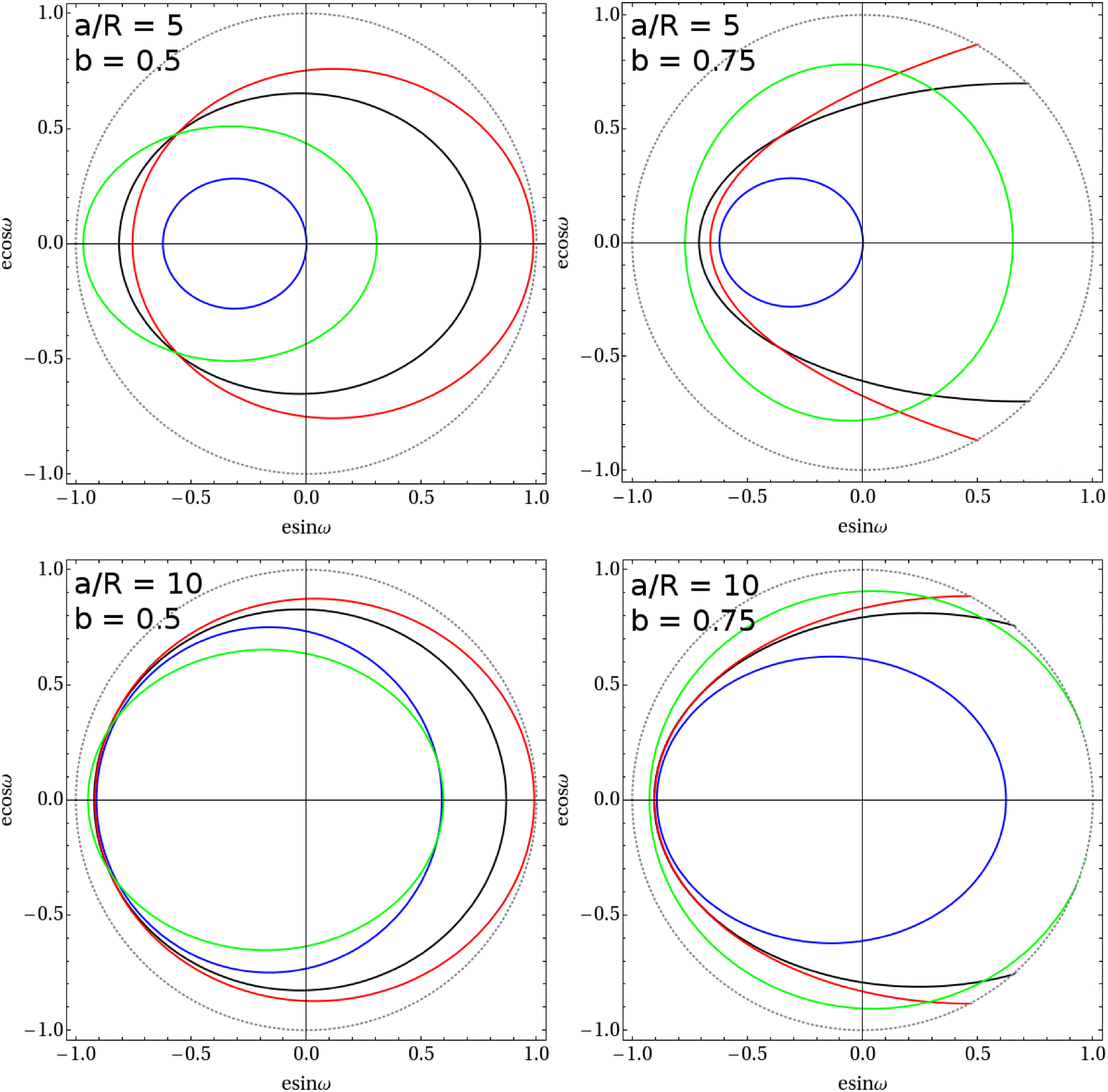}
\caption{\emph{Loci of points for which the accuracy is better than 99\% for all four candidate expressions, as a function of eccentricity. The $T_1$ expression offers both consistency and excellent accuracy. Other system parameters fixed to typical transit values. Blue is for $T_{\mathrm{TS05}}$, black is for $T_{\mathrm{W10}}$, red is for $T_1$ and green for $T_2$. The gray ellipse represents the allowed physical limits.}} \label{fig:fig2}
\end{center}
\end{figure*}

\subsection{Additional Tests}

We briefly discuss additional tests we performed for two sets of $10^7$ different hypothetical transiting exoplanet systems; one for eccentricities $0.0<e<1.0$ and one for $0.9<e<1.0$. In all cases, we randomly generated the system parameters weighted by the transit probability and calculated the deviation of the various formulae.

We found that the $T_1$ expressions was consistently the most accurate, with the W10 of similar accuracy but higher assymmetry. We therefore find that the results yield a overall preference for the $T_1$ approximation.  We note that authors using the $T_{\mathrm{W10}}$ formulation can also expect an extremely good approximation but for the remainder of this paper we will only consider using $T_1$ for the later derivations. We may define the `improvement' of the $T_1$ expression relative to the $T_{\mathrm{TS05}}$ equation as:

\begin{equation}
\mathcal{I}_{\mathrm{T1}} = [ (\mathcal{D}_{\mathrm{TS05}}/\mathcal{D}_{\mathrm{T1}} ) - 1 ]*100
\end{equation}

Where $\mathcal{I}$ is measured in \%.  We can see that if TS05 gives a lower deviation (i.e. more accurate solution), we will obtain $\mathcal{I} < 0$\% whereas if the candidate expression gives a closer solution we obtain $\mathcal{I} > 0$\% and is essentially the percentage improvement in accuracy.  For the range $0<e<1$, we find the mean value of this parameter is $\mathcal{I}_{\mathrm{T1}} = 210$\% and for the range $<0.9<e<1.0$ we find $\mathcal{I}_{\mathrm{T1}} = 458$\%. We note that one caveat of these tests is that they are sensitive to the a priori inputs.

For the case of Kepler photometry, following the method of \citet{kip09c}, we estimate that the typical measurement uncertainty on $T$ will be $\sim$0.1\% in most cases. We find that $T_1$ is accurate to 0.1\% or better over a range of $|e\sin\omega|<0.5$ and $|e\cos\omega|<0.85$ on average.

\section{Analytic Investigations}
\subsection{The Consequences of Using Circular Expressions for Eccentric Orbits}

SMO03 showed that the $1^{\mathrm{st}}$ to $4^{\mathrm{th}}$ contact duration, $t_T$, and the $2^{\mathrm{nd}}$ the $3^{\mathrm{rd}}$ contact duration, $t_F$, may be used to derive $a_R$, $i$, $b$ and the stellar density, $\rho_*$.  We here consider how biased these retrieved parameters would be if we used the circular equations for an eccentric orbit.  From here, we will employ the $T_{1}$ expression for the transit duration, as this equation has been shown to provide the greatest accuracy in the previous section.  According, to SMO03, the circular transit durations are given by:

\begin{align}
t_T(\mathrm{SMO03}) &= \frac{P}{\pi} \arcsin\Bigg(\frac{\sqrt{(1+p)^2 - a_R^2 \cos^2 i}}{a_R \sin i}\Bigg) \\
t_F(\mathrm{SMO03}) &= \frac{P}{\pi} \arcsin\Bigg(\frac{\sqrt{(1-p)^2 - a_R^2 \cos^2 i}}{a_R \sin i}\Bigg)
\end{align}

Modification of the $T_{1}$ solution gives:

\begin{align}
t_{T,1} &= \frac{P}{\pi} \frac{\varrho_c^2}{\sqrt{1-e^2}} \arcsin\Bigg(\frac{\sqrt{(1+p)^2 - a_R^2 \varrho_c^2 \cos^2 i}}{a_R \varrho_c \sin i}\Bigg) \\
t_{F,1} &= \frac{P}{\pi} \frac{\varrho_c^2}{\sqrt{1-e^2}} \arcsin\Bigg(\frac{\sqrt{(1-p)^2 - a_R^2 \varrho_c^2 \cos^2 i}}{a_R \varrho_c \sin i}\Bigg)
\end{align}

Using (28) and (29), SMO03 show that the impact parameter may be retrieved by using:

\begin{equation}
[b_{\mathrm{derived}}(\mathrm{SMO03})]^2 = \frac{(1-p)^2 - \frac{\sin^2(t_F \pi/P)}{\sin^2(t_T \pi/P)} (1+p)^2}{ 1-\frac{\sin^2(t_F \pi/P)}{\sin^2(t_T \pi/P)} }
\end{equation}

Using the same equations for an eccentric orbit gives:

\begin{align}
&[b_{\mathrm{derived}}(\mathrm{SMO03})]^2 = 1 + p^2 + 2p \cdot \nonumber \\
&\Bigg(\frac{ \sin^2[\frac{\varrho_c^2}{\sqrt{1-e^2}} \arcsin(\frac{\sqrt{(1-p)^2 - b^2}}{a_R \varrho_c \sin i})] + \sin^2[\frac{\varrho_c^2}{\sqrt{1-e^2}} \arcsin(\frac{\sqrt{(1+p)^2 - b^2}}{a_R \varrho_c \sin i})] }{ \sin^2[\frac{\varrho_c^2}{\sqrt{1-e^2}} \arcsin(\frac{\sqrt{(1-p)^2 - b^2}}{a_R \varrho_c \sin i})] - \sin^2[\frac{\varrho_c^2}{\sqrt{1-e^2}} \arcsin(\frac{\sqrt{(1+p)^2 - b^2}}{a_R \varrho_c \sin i})] }\Bigg)
\end{align}

Where it is understood that for terms on the right-hand side with $b$ in them, we are referring to the true impact parameter, $b = a_R \varrho_c \cos i$.  We plot this function in the case of $a_R = 10$, $b^2=0.5$ and $p=0.1$ in figure 3. Making small-angle approximations, this yields $b_{\mathrm{derived}}^2 \simeq b^2$.  However, for larger $e\sin\omega$ and $e\cos\omega$ values, the overall effect is to overestimate $b$ for eccentric orbits.

\begin{figure}
\begin{center}
\includegraphics[width=8.4 cm]{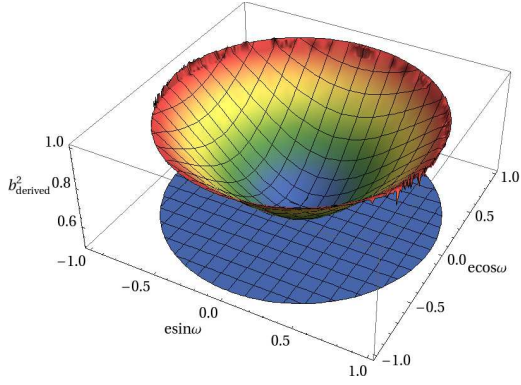}
\caption{\emph{If one uses the circular expressions, the retrieved impact parameter (squared) is heavily biased by eccentricity.  In this example, the true value of $b^2$ is 0.5 but the introduction of eccentricity causes $b^2$ to be overestimated.}} \label{fig:fig3}
\end{center}
\end{figure}

In addition to the impact parameter, SMO03 proposed that the parameter $a_R = a/R_*$ may be derived using:

\begin{equation}
[a_{R,\mathrm{derived}}(\mathrm{SMO03})]^2 = \frac{(1+p)^2 - b_{\mathrm{derived}}^2}{\sin^2 (t_T \pi/P)} + b_{\mathrm{derived}}^2
\end{equation}

If we use the assumption $b_{\mathrm{derived}} \simeq b$, then this equation yields:

\begin{align}
&[a_{R,\mathrm{derived}}(\mathrm{SMO03})]^2 = b^2 + \nonumber \\
&((1+p^2)-b^2) \csc^2\Bigg[\frac{\varrho_c^2}{\sqrt{1-e^2}} \arcsin\Big(\frac{\sqrt{(1+p)^2-b^2}}{a_R \varrho_c \sin i}\Big)\Bigg]
\end{align}

With small-angle approximations, we have:

\begin{equation}
a_{R,\mathrm{derived}} \simeq a_R \sqrt{\varrho_c^2 \cos^2 i + \frac{(1-e^2) \sin^2 i}{\varrho_c^2}}
\end{equation}

\begin{figure}
\begin{center}
\includegraphics[width=8.4 cm]{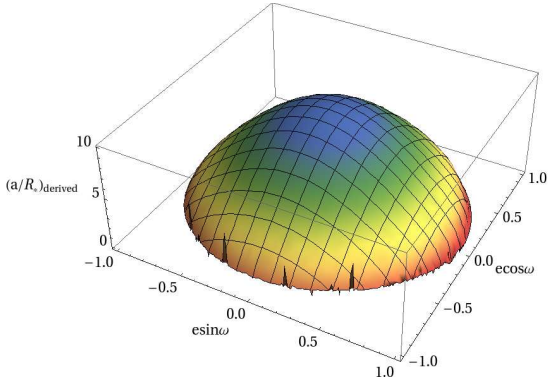}
\caption{\emph{If one uses the circular expressions, the retrieved value of $a/R_*$ is heavily biased by eccentricity.  In this example, the true value of $a/R_*$ is 10 but the introduction of eccentricity causes $a/R_*$ to be underestimated.}} \label{fig:fig4}
\end{center}
\end{figure}

The term inside the square root goes to unity for circular orbits, as expected.  The deviation in $a_R$ can be seen to become quite significant for eccentric orbits, as seen in figure 4 where the exact expression for (34) is plotted. This will have significant consequences for our next parameter, the stellar density.  Stellar density is related to $a_R$ by manipulation of Kepler's Laws:

\begin{align}
\rho_* + p^3 \rho_P = \frac{3 \pi}{G P^2} a_R^3 \nonumber \\
\rho_* \simeq \frac{3 \pi}{G P^2} a_R^3
\end{align}

Where the approximation is made using the assumption $p \ll 1$.  We can therefore see that:

\begin{align}
\rho_{*,\mathrm{derived}} &\simeq \rho_* \Bigg[\varrho_c^2 \cos^2 i + \frac{(1-e^2) \sin^2 i}{\varrho_c^2}\Bigg]^{3/2} \\
\rho_{*,\mathrm{derived}} &\simeq \rho_* \Psi = \Bigg[\frac{(1+e \sin \omega)^3}{(1-e^2)^{3/2}}\Bigg] \rho_*
\end{align}

Where in the second line we have assumed that $i \simeq \pi/2$.  A series expansion of $\Psi$ into first order of $e$ yields $\Psi \simeq 1 + 3 e \sin \omega + \mathcal{O}[e^2]$.  So observers neglecting an eccentricity of $e \sim 0.1$ may alter the stellar density by 30\%.  As an example, if we decreased the density of a solar type G2V star by 30\%, the biased average stellar density would be more consistent with a star of spectral type K0V. Indeed, asteroseismologically determined stellar densities of transiting systems could be used to infer $\Psi$.

\begin{figure}
\begin{center}
\includegraphics[width=8.4 cm]{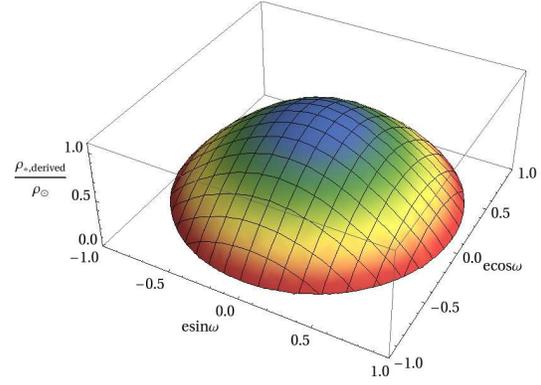}
\caption{\emph{If one uses the circular expressions, the retrieved value of $\rho_*$ is heavily biased by eccentricity.  In this example, the true value of $\rho_*$ is 1$\rho_{\odot}$ but the introduction of eccentricity causes $\rho_*$ to be underestimated.}} \label{fig:fig5}
\end{center}
\end{figure}

This density bias, which is plotted in figure 5, could be extremely crucial in the search for transiting planets.  Many discovery papers of new transiting planets have only sparse radial velocity data and usually no secondary eclipse measurement.  As a result, the uncertainity on the eccentricity is very large.

Critically, planets are often accepted or rejected as being genuine or not on the basis of this lightcurve derived stellar density.  If the lightcurve derived stellar density is very different from the combination of stellar evolution and spectroscopic determination, these candidates are generally regarded as unphysical.  This method of discriminating between genuine planets and blends, which may mimic such objects, was proposed by \citet{sea03} (see \S6.3 of SMO03) but crucially only for circular orbits.

Since the typical upper limit on $e$ is around 0.1 in discovery papers, then the lightcurve derived stellar density also has a maximum possible error of $\sim30$\%. In practice, the uncertainty on $e$ will result in a larger uncertainty in $\rho_*$.  Typical procedure is to fix $e=0$ if the radial velocity data is quite poor, despite the fact the upper limit on $e \sim 0.1$.  As a result, the posterior distribution of $\rho_*$ would be artificially narrow and erroneous if $e \neq 0$.  We propose that global fits should allow $e$ to vary when analyzing radial velocity and transit photometry, as well as a fixed $e=0$ fit for comparison.  This would allow the full range of possible eccentricities to be explored, which would result in a broader and more accurate distribution for $b$, $a_R$ and critically $\rho_*$.

\subsection{General Solution for the Physical Parameters}

In the previous subsection we saw how using the circular expressions to derive $a_R$ and $\rho_*$ can lead to severe errors for even mildly eccentric systems.  We here present expressions which will recover excellent approximations values for $b$, $a_R$ and $\rho_*$.  The new equations are given by:

\begin{align} 
b^2 &= \frac{(1-p)^2 - \frac{\sin^2[(t_F \pi \sqrt{1-e^2})/(P \varrho_c^2)]}{\sin^2[(t_T \pi \sqrt{1-e^2})/(P \varrho_c^2)]} (1+p)^2}{ 1-\frac{\sin^2[(t_F \pi \sqrt{1-e^2})/(P \varrho_c^2)]}{\sin^2[(t_T \pi \sqrt{1-e^2})/(P \varrho_c^2)]} } \\
a_R^2 &= \frac{(1+p^2) - b^2}{\varrho_c^2 \sin^2[(t_T \pi \sqrt{1-e^2})/(P \varrho_c^2)]} + \frac{b^2}{\varrho_c^2} \\
\rho_{*} &= \frac{3 \pi}{G P^2} a_R^3 - p^3 \rho_P
\end{align}

These expressions can be shown to reduce down to the original SMO03 derivations if $e \rightarrow 0$ (equations (7) \& (8) of SMO03).  The new stellar density parameter may be used with floating $e$ and $\omega$ values to correctly estimate the probability distribution of this critical parameter.

\subsection{The Transit `Width' Duration}

In the previous subsection, we have derived the physical parameters in terms of $t_T$ and $t_F$.  We may naively assume that this is interchangeable with expressions in terms of $T$ and $\tau$.  Assuming the transit lightcurve is symmetric (exactly valid for circular orbits and a very good approximation for eccentric orbits), the following relations between these two definitions exist:

\begin{align}
\tau &= \frac{t_T - t_F}{2} \\
T &\neq \frac{t_T + t_F}{2} = W
\end{align}

For the latter, $T = W$ for the trapezoid approximated lightcurve only.  This is because $T$ is defined as when the sky-projection of the planet's centre is touching the stellar limb i.e. $S=1$.  At this point, the fraction of the planetary disc occulting the stellar disc is not equal to one half of the total in-transit occulted area.  In contrast, we here define $W$ as the duration between the midway of the ingress to the midway of the egress.  We can intuitionally see at the moment $S=1$, less than half of the total area must be occulted and therefore $T > W = (t_T + t_F)/2$.

A further validation of this can be seen by writing down the equations for $t_T$ and $t_F$ and combining them using arcsin trigometric identities.  For the simple case of a circular orbit, the resultant expression would give:

\begin{align}
&W(e=0) = \frac{P}{\pi} \arcsin\Bigg(\frac{\sqrt{1-\sqrt{1-\alpha^2}}}{\sqrt{2}}\Bigg) \\
&\alpha = \Big(a_R \sin i\Big)^{-1} \Bigg[[(1+p)^2 - b^2]^{1/2} [1- \nonumber \\
&(1-p^2) + b^2]^{1/2} - [(1-p)^2 - b^2]^{1/2} [1- (1+p^2) + b^2]^{1/2}\Bigg]
\end{align}

It may be shown that $\alpha \neq \sqrt{1-b^2}/(a_R \sin i)$ and thus $W \neq T$.  In the same manner as we derived $W$, $\tau$ may also be written as a combination of the relevant arcsin functions.  However, we can already see that such an expression will also be extremely laborius.  This means that writing down the expressions for $b(T,\tau)$ and $a_R(T,\tau)$ is much more challenging than that for $b(t_T,t_F)$ and $a_R(t_T,t_F)$ and we were unable to find an exact inversion relation.  Indeed, finding exact expressions for these inverse relations is unncessary as we may retrieve $b$ and $a_R$ using equations (40) and (41).

For a circular orbit and in the limit $a_R \gg 1$, the relative difference betwen $T$ and $W$ may be written as:

\begin{equation}
\frac{T_1-W_1}{T_1} \simeq \frac{2 \sqrt{1-b^2}-\sqrt{(1+p)^2-b^2} - \sqrt{(1-p)^2 - b^2}}{2 \sqrt{1-b^2}}
\end{equation}

Notice how for $p\rightarrow0$ this expression yields zero, which is expected since the planet now takes infintessimal size.  The denominator also reveals that the difference diverges rapidly for near-grazing transits i.e. $b \rightarrow 1$.

\section{Applications to Lightcurve Fitting}
\subsection{Fitting Parameter Sets: \{$t_c$, $p^2$, $W_1$, $\tau_1$\}}

The transit lightcurve is essentially described by four physical parameters, which form the parameter set \{$t_c$, $p$, $a_R$, $b$\}.  However, efforts to fit transit lightcurves using this parameter set is known to be highly inefficient due to large inter-parameter correlations, in particular between $a_R$ and $b$.  \citet{car08} used a Fisher analysis to show that for a symmetric lightcurve, which is approximated as a piece-wise linear model (i.e. a trapezoid), a superior parameter set to fit for is given by \{$t_c$, $p^2$, $T$, $\tau$\} where $\tau$ is the ingress or equivalently egress duration assuming a symmetric lightcurve.  In our case, these parameters become \{$t_c$, $p^2$, $T_1$, $\tau_1$\}.  The authors reported that using this parameter set decreased the correlation lengths in a Markov Chain Monte Carlo (MCMC) fit by a factor of $\sim 150$.

In the \citet{car08} analysis, the lightcurve is a symmetric trapezoid and therefore $W = T$.  As we have already seen, this is not true for a real lightcurve.  This raises the ambiguity as to whether this fitting parameter should be $W$ or $T$ for real lightcurves.

One advantage of the $T$ parameter is that it is independent of $p$, whereas $W$ is not.  With one degree less of freedom than $W$, $T$ will always exhibit lower correlations and may be determined to lower uncertainity.  This makes $T$ ideal for transit duration variations (TDV) studies, as pointed out by \citet{kip09c}.  However, as we saw in \S5.3, there presently exists no known expression for converting $T$ and $\tau$ into the physical parameter set which is used to actually generate a model lightcurve.

When fitting a transit lightcurve, our hypothetical algorithm must make trial guesses for `the fitting parameter set' which is then mapped into `the physical parameter set'.  These physical parameters are then fed into a transit lightcurve model generator, allowing for the goodness-of-fit between the trial model and the observations to be made.  This mapping proceedure is unavoidable since the transit lightcurve is essentially generated by feeding the sky-projected planet-star separation, $S$, as a function of time, into a lightcurve generating code like that of \citet{man02}.  Since $S$ (equation 4) is a function of the physical parameter set, and not the fitting parameter set, the mapping between the two sets is a pre-requisite for any lightcurve fitting algorithm.

Unless an approximation is made that $T_1 \simeq W_1$, there currently exists no expressions which perform this mapping proceedure for the fitting parameter set \{$t_c$, $p^2$, $T_1$, $\tau_1$\}\footnote{Although mappings have been proposed, they all make the assumption $T=W$}.  Specifically, there currently exists no exact expression for $b(T_1,\tau_1,p)$ and $a_R(T_1,\tau_1,p)$.  Therefore, \{$t_c$, $p^2$, $T_1$, $\tau_1$\} cannot be used as a fitting parameter set unless we assume $T_1 \simeq W_1$ and use \{$t_c$, $p^2$, $W_1$, $\tau_1$\}

Fortunately, the consequences of making this assumption will not be severe, in most cases.  This is because the trial fitting parameters serve only one function - to produce trial physical parameters.  These trial physical parameters may be slightly offset from the exact mapping but this is not particularly crippling since the model lightcurve is still generated exactly based upon these trial physical parameters.  The only negative consequence of using this method is that an additional correlation has been introduced into the fitting algorithm since the offset between $T$ and $W$ will be a function of $b$ and $p$.  This correlation will be largest for near-grazing transits since equation (47) tends towards $\infty$ as $b\rightarrow 1$.  In general, we wish to avoid such correlations as much as possible to allow the algorithm to most efficiently explore the parameter space.

\subsection{Fitting Parameter Sets: \{$t_c$, $G_1$, $W_1$, $A_1$\}}

For a trapezoid approximated lightcurve, \citet{car08} showed that the fitting parameter correlations are decreased further by using the parameter set \{$t_c$, $G$, $W$, $A$\}, where $A$ is the area of the trapezoid-approximated lightcurve, and $G$ is the gradient of the ingress/egress (note we have changed the original notation from $S$ to $G$ to avoid confusion with equation 4)\footnote{We note that \citet{car08} proposed an additional slightly improved parameter-set, but this set required reliable prior estimates of $b$ and lacked a physical interpretation}.

The area of the trapezoid lightcurve is given by $\delta(t_T + t_F)/2$ where $\delta = p^2$.  The gradient of a trapezoid slope is given by $\delta/\tau$.  Since $\tau = (t_T - t_F)/2$ then both $A$ and $G$ may be written as a function of $t_T$ and $t_F$ only, thus obviating the use of $T$ and the associated issues discussed in the previous subsection.

In order to proceed, a mapping from \{$G$, $W$, $A$\} $\rightarrow$ \{$p$, $a_R$, $b$\} is required for accomplishing this goal.  The exact solutions for $G$, $W$ \& $A$ may be found by solving the quartic equation discussed in \S2.  However, the roots of this equation yields $G(p,a_R,b)$, $W(p,a_R,b)$ \& $A(p,a_R,b)$ whereas we need the inverse relations.  Since no concise analytic solution for the inverse relations currently exists, these inverse relations would have to be calculated through a numerical iteration but such a process would need to be repeated for every single trial leading to vastly greater computation time for a fitting algorithm.

Therefore, a practical compromise is to use approximate formulae $G_1(p,a_R,b)$, $W_1(p,a_R,b)$ \& $A_1(p,a_R,b)$, which are easily manipulated to give the inverse relations: $p(G_1,W_1,A_1)$, $a_R(G_1,W_1,A_1)$ and $b(G_1,W_1,A_1)$.

It is critical to understand that using equations for a circular orbit or an approximate eccentric expression of poor accuracy will cause fitting algorithms to wander into unphysical solutions and/or increase inter-parameter correlations for planets which are eccentric, near-grazing, very close-in, etc.  It is therefore imperative to use a mapping which is as accurate as possible in order to have a robust fitting algorithm.  The mappings to convert the trial \{$G_1$, $W_1$, $A_1$\} into the physical parameters \{$p$, $a_R$, $b$\} are given by equations (40) and (41) combined with the following replacements:

\begin{align}
p &= \sqrt{\frac{A_1}{W_1}} \\
t_{T,1} &= W_1 + \frac{A_1}{W_1 G_1} \\
t_{F,1} &= W_1 - \frac{A_1}{W_1 G_1}
\end{align}

We note that the favoured parameter set derived by \citet{car08} assumed a symmetric lightcurve which is not strictly true for $e > 0$.  However, for an eccentric orbit, the degree of assymetry between the ingress and egress is known to be very small (K08, W10) and thus may be neglected for the purposes of choosing an ideal fitting parameter set.

\subsection{Fitting Parameter Sets: \{$t_c$, $p^2$, $\zeta/R_*$, $b^2$\}}

The two parameter sets proposed so far have both included $W_1$.  Since we know $W_1$ is a function of $p$ but $T_1$ is not, any parameter set using $W_1$ will likely exhibit larger correlations since there is an extra parameter dependency.  However, a parameter set based upon $T_1$ would have to satisfy the condition that it can be inverse mapped into the physical parameters.

A search through the literature finds just such a parameter set. \citet{bak07} proposed the fitting parameter set \{$t_c$, $p^2$, $\zeta/R_*$, $b^2$\}, where $\zeta/R_*$ is defined by:

\begin{align}
\frac{\zeta}{R_*} &= \frac{2}{T_{\mathrm{TS05}}} \nonumber \\
\frac{\zeta}{R_*} &= \frac{2 \pi a}{P R_*} \frac{1+e \sin \omega}{\sqrt{1-e^2}\sqrt{1-b^2}}
\end{align}

$\zeta/R_*$, originally defined by \citet{mur99}, can be seen to be reciprocal of one half of the transit duration as defined by TS05.  Unlike the \{$t_c$, $p^2$, $T$, $\tau$\} parameter set, we do not need to assume $T = W$ to produce an inverse mapping.  By using $T_{\mathrm{TS05}}$ and $b^2$, an exact inverse mapping to the physical parameters is possible which offers significant advantages.

Having satisfied the criteria of being both a decorrelated parameter and inverseable, $\zeta/R_*$ would appear to an excellent candidate for lightcurve fitting.  A further improvement to this parameter is possible by using the new $T_1$ approximation for the transit duration.  Let us define:

\begin{align}
\frac{\Upsilon}{R_*} &= \frac{2}{T_1} \nonumber \\
\frac{\Upsilon}{R_*} &= \frac{2\pi}{P} \frac{\sqrt{1-e^2}}{\varrho_c^2} \Bigg[\arcsin\Bigg(\frac{\sqrt{1-a_R^2 \varrho_c^2 \cos^2i}}{a_R \varrho_c \sin i}\Bigg)\Bigg]^{-1}
\end{align}

The inverse mapping would use the expression:

\begin{equation}
a_R^2 = \frac{1-b^2}{\varrho_c^2} \csc^2\Big[\frac{2 \pi \sqrt{1-e^2}}{P \varrho_c^2 (\Upsilon/R_*)}\Big] + \frac{b^2}{\varrho_c^2}
\end{equation}

\subsection{Circular Orbit Example}

Despite the analytic arguments made so far, the clearest validation of which fitting parameter set to employ may be resolved through numerical simulations.  This may be done by considering an example system, generating a lightcurve, adding noise and then refitting using an MCMC routine which outputs the inter-parameter correlations.  We note that analytic expressions for the covariances may be found through Fisher information analysis through the calculation of the relevant partial derivatives \citep{pal08}. However, the equations describing the lightcurve, as given by \citet{man02} are quite elaborate and such an analysis remains outside of the scope of this paper. Currently, there exists no exact Fisher information analysis in the literature to draw upon. \citet{car08} avoided this problem by making a trapezoid approximation of the lightcurve and then implementing a Fisher analysis. As discussed earlier, this requires that we assume $T=W$, which in itself introduces a host of correlations which would be missed by the Fisher analysis methodology. Therefore, exact numerical testing provides a useful alternative to avoid these issues.

First, we consider a super-hot Jupiter on a circular orbit with a planet-star separation of $a_R =3.5$ from a Sun-like star ($P=0.76$days).  We choose to consider a near-grazing transit with $b=0.9$ corresponding to an orbital inclination of $75.1^{\circ}$.  The lightcurve is generated using the \citet{man02} algorithm with no limb-darkening and $0.25$mmag Gaussian noise over a 30 second cadence.  The lightcurve is then passed onto a MCMC fitting algorithm where we try several different parameter sets:

\begin{itemize}
\item \{$t_c$, $p$, $a_R$, $b$\}: the physical parameter set.
\item \{$t_c$, $p^2$, $\zeta/R_*$, $b^2$\}: a suggested lightcurve fitting parameter by \citet{bak07}, based upon the TS05 duration expressions.
\item \{$t_c$, $p^2$, $\Upsilon/R_*$, $b^2$\}: a modified form of the fitting parameter by \citet{bak07}, accounting for the improved approximate expression for $T$.
\item \{$t_c$, $p^2$, $W_1$, $\tau_1$\}: a suggested set by \citet{car08}, where $W_1$ and $\tau_1$ are calculated using the expressions presented in this paper.
\item \{$t_c$, $G_1$, $W_1$, $A_1$\}: a second suggested set by \citet{car08}, where $G_1$, $W_1$ and $A_1$ are calculated using the expressions presented in this paper.
\end{itemize}

In the MCMC runs, we set the jump sizes to be equal to $\sim 1$-$\sigma$ uncertainties from a preliminary short-run.  We then start the MCMC from 5-$\sigma$'s away from the solution for each parameter, and use 500,000 trials with a 100,000 burn-in time.  We then compute the cross-correlations between the various parameters in trials which are within $\Delta \chi^2 = 1$ of $\chi^2|_{\mathrm{best}}$ (errors rescaled such that $\chi^2|_{\mathrm{best}}$ equals number of data points minus the degrees of freedom).  We calculate the inter-parameter correlations and construct correlation matrices for each parameter fitting set.  As an example, the correlation matrix for the $\{t_c, p, a_R, b\}$ parameter set is given by:

\begin{eqnarray} \nonumber
&\mathrm{Corr}(\{t_c, p, a_R, b\},\{t_c, p, a_R, b\})= \nonumber \\
& \left(
\begin{array}{cccc}
         1              & \mathrm{Corr}(t_c,p) & \mathrm{Corr}(t_c,a_R) & \mathrm{Corr}(t_c,b) \\
 \mathrm{Corr}(p,t_c)   &            1         & \mathrm{Corr}(p,a_R)   & \mathrm{Corr}(p,b) \\
 \mathrm{Corr}(a_R,t_c) & \mathrm{Corr}(a_R,p) &             1          & \mathrm{Corr}(a_R,b) \\
 \mathrm{Corr}(b,t_c)   & \mathrm{Corr}(b,p)   & \mathrm{Corr}(p,a_R)   &           1
\end{array} \nonumber
\right)
\end{eqnarray}

We then calculate the semi-principal axes of correlation ellipsoid by diagnolizing the matrices.  For a completely optimal parameter set, this diagnolized matrix would be the identity matrix.  We quantify the departure of each proposed parameter set from the identity matrix by calculating $\sum_{i=1}^4 |M_{i,i}-1|$ where $M$ is the diagnolized correlation matrix.  We display the results in upper half of table 1.

\begin{table}
\caption{\emph{For each proposed lightcurve fitting parameter set (left column), we calculate the inter-parameter correlation matrices in the examples of i) a hypothetical near-grazing hot-Jupiter on a circular orbit ii) a system similar to the eccentric planet HD 80606b.  We diagnolize the correlation matrices to give $M$ and then quantify the departure from a perfectly optimal parameter set (right column), where it is understood that 0 corresponds to optimal and larger values correspond to greater inter-parameter correlations.}} 
\centering 
\begin{tabular}{c c} 
\hline\hline 
Parameter Set & $\sum_{i=1}^4 |M_{i,i}-1|$ \\ [0.5ex] 
\hline
Circular orbit example \\
\{$t_c$, $p$, $a_R$, $b$\} & 2.19333 \\
\{$t_c$, $p^2$, $\zeta/R_*$, $b^2$\} & 1.71236 \\
\{$t_c$, $p^2$, $\Upsilon/R_*$, $b^2$\} & 1.32974 \\
\{$t_c$, $p^2$, $W_1$, $\tau_1$\} & 1.67485 \\
\{$t_c$, $G_1$, $W_1$, $A_1$\} & 2.23820 \\
\hline
Eccentric orbit example \\
\{$t_c$, $p$, $a_R$, $b$\} & 2.46676 \\
\{$t_c$, $p^2$, $\zeta/R_*$, $b^2$\} & 1.56816 \\
\{$t_c$, $p^2$, $\Upsilon/R_*$, $b^2$\} & 1.56776 \\
\{$t_c$, $p^2$, $W_1$, $\tau_1$\} & 1.57730 \\
\{$t_c$, $G_1$, $W_1$, $A_1$\} & 2.52948 \\
\hline
\end{tabular}
\label{table:nonlin} 
\end{table}

The correlations of the physical parameter set are predictably very large, in particular between $a_R$ and $b$ which approaches unity.  An inspection of the correlations for the other proposed parameter sets suggests that the modified \citet{bak07} formulation offers the lowest correlations.  The effect of modifying $\zeta/R_*$ to $\Upsilon/R_*$ produces a clear improvement in the corresponding correlations, as seen in figure 6.  As a result, the numerical tests support using the modified form of the \citet{bak07} parameter set.

\begin{figure}
\begin{center}
\includegraphics[width=8.4 cm]{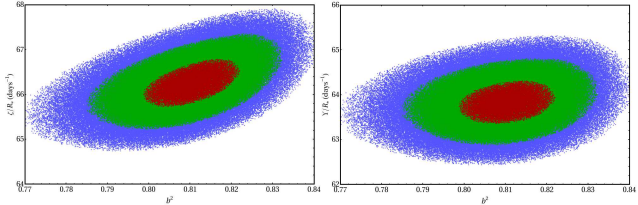}
\caption{\emph{Comparison of the correlations between $\zeta/R_*$ against $b^2$ and $\Upsilon/R_*$ against $b^2$.  Data comes from fitting a synthetic hot-Jupiter lightcurve on a circular, near-grazing orbit with an MCMC algorithm.  The new $\Upsilon/R_*$ parameter provides two-fold lower correlation and preserves the ability to be inversely mapped to more physical parameters.  The three different types of shading represent the 1-$\sigma$, 2-$\sigma$ and 3-$\sigma$ confidence regions.}} \label{fig:fig6}
\end{center}
\end{figure}

\subsection{Eccentric Orbit Example}

As a second example, we consider a highly eccentric orbit.  In this case, we choose to use the real system HD 80606b.  HD 80606b is highly eccentric planet with $e\sim0.93$ first discovered through radial velocity (Naef et al. 2001) and then later found to go through both secondary and primary eclipse (Laughlin et al. 2005; Fossey et al. 2009).  Using the system parameters from \citet{win09}, we generate a synthetic lightcurve of cadence 1 minute and 1.0 mmag Gaussian noise.  We ensure the quantity of out-of-transit baseline data is approximately equal to the full transit duration.  We adopt the same methodology as before to calculate the inter-parameter correlations of the various sets with the results shown in the lower half of table 1.

In these tests, we find three sets produce approximately the same optimization but the lowest correlations occur for the modified \citet{bak07} set again.  The difference between the correlations in the unmodified and the modified \citet{bak07} parameter set is extremely small, but there is a very slight improvement in the modified version.  The difference for circular orbits was larger due to the more grazing transit and the fact the TS05 expressions do not reduce down to the circular form.  In conclusion, we advocate using the modified \citet{bak07} parameter set i.e. $\{t_c, p^2, \Upsilon/R_*, b^2\}$ to most efficiently fit transit lightcurves.

\section{Secular Transit Duration Variation}

Transit duration variation (TDV) can occur two possible formats: i) periodic change ii) secular change.  As an example, periodic change in the transit duration was predicted to occur for a transiting planet with a companion moon by Kipping (2009a, b).  Conceptually analagous to the radial velocity of finding planets, the sky-projected tangential velocity of the planet oscillates around some local mean value as a result of the moon's gravitational tug.  These changes in tangential velocity induce variations in the transit duration, allowing for the detection of exomoons as small as 0.2$M_{\oplus}$ with space-based photometry (Kipping et al. 2009c).

Secular changes in transit duration can be caused by numerous possible scenarios.  \citet{jor08} showed that apsidal precession would induce changes in $T$ and used the TS05 equation to predict the size of these changes.  As we have already demonstrated a better formulation for $T$ is possible; we will here present an improved equation for the rate of change in $T$ due to apsidal precession, or essentially changes in $\omega$.

\citet{jor08} argued that apsidal precession can be caused by stellar oblateness, general relativistic effects and/or a perturbing planet.  Additionally, \citet{mur99} showed that in general nodal precession should also occur whenever apsidal precession occurs, leading to changes in the orbital inclination angle, $i$.  These changes lead to another form of secular TDV.

Additionally, we consider here that falling planets, such as proposed for WASP-18b (Hellier et al. 2009), would experience a changing semi-major axis, $a$, leading to another form of secular TDV.  Finally, we will consider the effect of varying the orbital eccenticity.  All four possible TDVs will be derived here using $T_1$, since this expression demonstrated the greatest accuracy in numerical tests.

\subsection{Apsidal Precession}

Apsidal precession is the precession of the argument of periapse over time and it may be induced from several different effects including:

\begin{itemize}
\item General relativistic effects (Einstein 1915; P\'{a}l \& Kocsis 2008)
\item Rotational quadrupole bulges on the planet (Sterne 1939)
\item Tides raised on the planet and the star (Sterne 1939)
\item Stellar quadrupole moment (Murray \& Dermott 1999)
\item Kozai mechanism (Kozai 1962)
\item Perturbing planets (Murray \& Dermott 1999; Miralda-Escud\'{e} 2002; Heyl \& Gladman 2007)
\end{itemize}

%

Out of these examples, planets on nearby orbits of masses $\geq M_{\oplus}$ are expected to produce the largest effect.  Thus the detection of apsidal precession could actually be used to infer the presence of Earth-mass planets.

As \citet{jor08} noted, apsidal precession should cause a change in the transit duration and in order to estimate the magnitude of this effect, the authors differentiated $T_{\mathrm{TS05}}$ with respect to $\omega$.  Having shown the $T_{1}$ offers substantial improvement over the $T_{\mathrm{TS05}}$ in the previous section, we are here able to provide an improved estimate for the secular TDV caused by apsidal precession:

\begin{equation}
\frac{\partial T}{\partial\omega} = \frac{P}{\pi} \frac{e \varrho_c^3 \cos \omega}{(1-e^2)^{3/2}} \Bigg(\frac{1}{\sqrt{1-b^2} \sqrt{a_R^2 \varrho_c^2 - 1}} - 2 \arcsin\Big(\frac{\sqrt{1-b^2}}{a_R \varrho_c \sin i}\Big) \Bigg)
\end{equation}

We can see that there are two terms counter-acting in our derived quantity.  The two terms can be understood to originate from the planet-star separation changing as a result of the precession which has two effects 1) decreasing the planet-star separation causes a near-grazing transit's impact parameter to decrease and thus increases $T$ (the first term) 2) decreasing the planet-star separation causes the tangential orbital velocity to increase and thus decreases $T$ (the second term).  The $\cos \omega$ term outside of the brackets determines the sign of which term causes an increase and which to decrease.

\citet{kop59} showed that the two effects approximately cancel out for $b \simeq 1/\sqrt{2}$.  The \citet{kop59} derivation is quite different for the ones produced in this paper.  It is done by first solving for the mid-eclipse time by a series expansion of the differential of the planet-star separation with respect to true anomaly, disregarding terms in $\sin^3i$ or higher.  The duration between the primary and secondary occultation is then solved for in another series expansion in first order of $e$.  Nevertheless, setting $b$ to this value, the terms inside the bracket of equation (54) become:

\begin{equation}
\frac{ \sqrt{2} }{ \sqrt{a_R^2 \varrho_c^2 - 1} } - 2 \arcsin\Bigg( \frac{1}{\sqrt{2} \sqrt{a_R^2 \varrho_c^2 - \frac{1}{2}}} \Bigg)
\end{equation}

Under the condition $a \gg R_*$, we find that equation (54) gives $\partial T/\partial\omega = 0$, in agreement with \citet{kop59}. For very close-in orbits, this does not appear to hold.

We may compare our estimate of the apsidal precession to equation (15) of \citet{jor08} , which was found by differentiating the expression of TS05 with respect to $\omega$.  The difference between the two expression is typically less than 1\% across a broad range of parameters.  However, if $b\simeq1/\sqrt{2}$, the difference between the two diverges and can reach 10\%-100\%.  Given the sensitivity of both equations to this critical value of $b$, we recommend numerical calculations over analytic approximations if $b$ is known to be close to 0.707.

%
%

\subsection{Nodal Precession}

Nodal precession causes changes in the orbital inclination of the planetary orbit, which would be a source of secular TDV.  The secular theory of \citet{mur99} predicts the rate of inclination change due to a perturbing planet as the nodes precess:

\begin{equation}
\frac{\partial i}{\partial t} = - \frac{\partial\omega}{\partial t} \Delta \Omega_{\mathrm{sky}}
\end{equation}

Where $\Delta \Omega_{\mathrm{sky}}$ is the ascending node of the perturbing planet relative to the ascending node of the transiting planet, measured clockwise on the plane of the sky.  Thus any occurence of apsidal precession from a perturbing planet will, in general, be coupled with nodal precession.  We may derive the rate of secular TDV from inclination change as before and find:

\begin{equation}
\frac{\partial T}{\partial i} = \frac{P}{\pi} \frac{\varrho_c^2 \sqrt{a_R^2 \varrho_c^2 - 1}}{\tan i \sqrt{1-e^2} \sqrt{1-b^2}}
\end{equation}

This expression has only one term and therefore we can see that decreasing the inclination towards a more grazing transit always yields a shorter transit duration, and vise versa.

\subsection{Falling Exoplanets}

Planetary bodies experience infall towards the host star through tidal dissipation and to a much lesser degree gravitational radiation.  The effects increase as the orbit becomes smaller leading to runaway fall-in.  Therefore, for very close-in exoplanets, the change in semi-major axis may be detectable.  The transit duration will vary as:

\begin{align}
&\frac{\partial T}{\partial a} = \frac{P}{\pi} \frac{\varrho_c^2}{a \sqrt{1-e^2}} \Bigg(\frac{3}{2} \arcsin\Big(\frac{\sqrt{1-b^2}}{a_R \varrho_c \sin i}\Big) - \nonumber \\
& \frac{1}{\sqrt{1-b^2} \sqrt{a_R^2 \varrho_c^2 - 1}}\Bigg)
\end{align}

As for apsidal precession, there are two countering components which are the same as before except for a slightly different constant in front of the first term.  This different constant means that the impact parameter at which both effects cancel has now changed to $b \simeq 1/\sqrt{3} = 0.577$.  This result could not be found in the previous literature and is of particular interest given the recent discovery of exoplanets on periods of around a day or less, for example WASP-18b (Hellier et al. 2009) with period of 0.94 days and $b = 0.25$.

\subsection{Eccentricity Variation}

Irregular satellites are known to exchange orbital inclination and eccentricity through the Kozai mechanism, which roughly conserves the value $\cos I \sqrt{1-e^2}$, where $I$ is the angle to the ecliptic.  Changes in orbital eccentricity are predicted to lead to long-term transit time variations (L-TTV) by \citet{kip08}, but here we consider the effect on the transit duration too.

\begin{align}
&\frac{\partial T}{\partial e} = \frac{P}{\pi} \frac{\varrho^3}{(1-e^2)^{5/2}} \Bigg[\frac{2 e+(1+e^2) \sin \omega}{\sqrt{1-b^2} \sqrt{a_R^2 \varrho_c^2 -1}} \nonumber \\
& - [3e+(2+e^2)\sin\omega] \arcsin\Big(\frac{\sqrt{1-b^2}}{a_R \varrho_c \sin i}\Big)\Bigg]
\end{align}

The two terms here seem to exhibit a more complicated inter-dependency which is physically based on the same idea of varying the planet-star separation.  The balance-point between the two effects occurs for:

\begin{equation}
b \simeq \sqrt{\frac{e+\sin\varpi}{(3e+(2+e^2) \sin\varpi}}
\end{equation}


\section{Conclusions}

We have derived and tested a new approximate expression for the transit duration of an extrasolar planet with non-zero orbital eccentricity (equation 15). The expression has been shown to analytically reduce down to the exact expressions for a circular orbit, unlike the most previously utilized equation.  In numerical tests, the new equation is shown to be more accurate than the other candidate expressions considered in this work, in particular for highly eccentric systems.  Quantitvely, the new expression yields a $>200$\% improvement in accuracy over the previously most utilized expression.

Manipulation of the new expression provides for a new lightcurve fitting parameter set which is based upon a modification of a previously proposed set by \citet{bak07}.  The new parameter set is shown to demonstrate the lowest mutual correlations compared to other most commonly used parameter sets and therefore yields the most efficient algorithm for fitting lightcurves.

Additionally, we have shown that the effect of even mild eccentricity can cause very large biases in the lightcurve derived stellar density, which is often used a method for discriminating between planets and blends in transit surveys.  Consequently, planetary candidates can be either falsely rejected or accepted for systems with poor contraints on the eccentricity.

Finally, we have used the new equation to derive the rates of secular transit duration variation (TDV) as a result of apsidal precession, nodal precession (e.g. due to a perturbing planet), in-falling extrasolar planets and eccentricity variation (e.g. Kozai mechanism).  These derivatives will provide for a more accurate interpretation of secular TDV.

\section*{Acknowledgments}

D. M. K. has been supported by HAT-NET \& HAT-South, the Harvard-Smithsonsian Center for Astrophysics predoctoral fellowships and the Science Technology and Facilities Council (STFC) studentships.  We are grateful to A. P\'{a}l for extremely helpful comments which improved the quality of this manuscript. Special thanks to G. Bakos for thought-provoking discussions on the subject. Thanks to G. Bakos and G. Tinetti for their continued support and advise.

\bsp

\appendix

\begin{table*}
\caption{\emph{List of important parameters used in this paper.}} 
\centering 
\begin{tabular}{l l l} 
\hline\hline 
Parameter & Name & Defintion \\ [0.5ex] 
\hline 
$S$ & Sky-projected separation & Sky-projected separation of the companion's centre \\
& & and the host star's centre in units of stellar radii \\
$R_*$ & Radius of the host star & Radius of the host star \\
$R_P$ & Radius of the companion & Radius of the companion \\ 
$p$ & Ratio-of-radii & Ratio of the companion's radius to the stellar radius ($R_P/R_*$) \\
$\delta$ & Geometric depth & The observed transit depth in the absence of limb darkening and \\
& & blended contamination, equal to $p^2$. \\
$t_{I}$ & First contact & Instant when $S = 1+p$ and $\mathrm{d}S/\mathrm{d}t < 0$ \\
$t_{II}$ & Second contact & Instant when $S = 1-p$ and $\mathrm{d}S/\mathrm{d}t < 0$ \\
$t_{C}$ & Mid-transit time & Instant when $\mathrm{d}S/\mathrm{d}t = 0$ i.e. inferior conjunction \\
$t_{III}$ & Third contact & Instant when $S = 1-p$ and $\mathrm{d}S/\mathrm{d}t > 0$ \\
$t_{IV}$ & Fourth contact & Instant when $S = 1+p$ and $\mathrm{d}S/\mathrm{d}t > 0$ \\
$t_T$ & Total duration & Time for companion to move between contact points I and IV \\
$t_F$ & Full duration & Time for companion to move between contact points II and III \\
$T$ & Transit duration & Time for companion to move across the stellar disc with entry \\ 
& & and exit conditions defined as $S=1$ \\
$W$ & Transit width & Mean value of $t_T$ and $t_F$ \\
$t_{12}$ & Ingress duration & Time for companion to move between contact points 1 and 2 \\
$t_{34}$ & Egress duration & Time for companion to move between contact points 3 and 4 \\
$\tau$ & Ingress/Egress duration & For circular orbits, $t_{12} = t_{34} = \tau$ \\
$T_1$ & $T_1$ duration & A one-term expression for $T$ derived in this work \\
$T_2$ & $T_2$ duration & A two-term expression for $T$ derived in this work \\
$T_{\mathrm{TS05}}$ & TS05 duration & Expression for $T$ derived by \citet{tin05} \\
$T_{\mathrm{W10}}$ & W10 duration & Expression for $T$ derived by \citet{win10} \\
$T_{\mathrm{SMO03}}$ & SMO03 duration & Expression for $T$ derived by \citet{sea03} \\
$f$ & True anomaly & True anomaly of the companion during its orbit around the host star \\
$E$ & Eccentric anomaly & Eccentric anomaly of the companion during its orbit around the host star \\
$M$ & Mean anomaly & Mean anomaly of the companion during its orbit around the host star \\
$\mu$ & Reduced mass & Reduced mass of the companion-star system \\
$J$ & Angular momentum & Angular momentum of the companion  \\
$a$ & Semi-major axis & Semi-major axis of the companion's orbit \\
$a_R$ & Semi-major axis & Semi-major axis of the companion's orbit in units of stellar radii \\
$e$ & Eccentricity & Orbital eccentricity of the companion's orbit \\
$\Omega$ & Longitude of the ascending node & Longitude of the ascending node of the companion's orbit \\
$\varpi$ & Longitude of pericentre & Longitude of pericentre of the companion's orbit \\
$\omega$ & Argument of pericentre & Argument of pericentre of the companion's orbit ($\omega = \varpi - \Omega$) \\
$\varrho$ & Companion-star separation & Companion-star separation in units of stellar radii  \\
$\varrho_c$ & Mid companion-star separation & Companion-star separation in units of stellar radii at the moment of mid-transit  \\
$P$ & Period & Orbital period of the companion \\
$b$ & Impact parameter & Value of $S$ when $\mathrm{d}S/\mathrm{d}t = 0$ \\
$i$ & Inclination & Orbital inclination of the companion's orbit relative to the line-of-sight of the observer \\
$D$ & Duration function & A parameter defined by \citet{kip08} \\
$f_a$ & True anomaly `a' & The true anomaly of the companion when $S=1$ and $\mathrm{d}S/\mathrm{d}t < 0$ \\
$f_b$ & True anomaly `b' & The true anomaly of the companion when $S=1$ and $\mathrm{d}S/\mathrm{d}t > 0$ \\
$f_c$ & Mid-transit true anomaly & The true anomaly of the companion when $\mathrm{d}S/\mathrm{d}t = 0$ \\
$f_M$ & Mean transit true anomaly & Mean of $f_b$ and $f_a$ \\
$f_H$ & Half true anomalistic duration & One half of the difference between $f_b$ and $f_a$ \\
$E_a$ & Eccentric anomaly `a' & The eccentric anomaly of the companion when $S=1$ and $\mathrm{d}S/\mathrm{d}t < 0$ \\
$E_b$ & Eccentric anomaly `b' & The eccentric anomaly of the companion when $S=1$ and $\mathrm{d}S/\mathrm{d}t > 0$ \\
$E_M$ & Mean transit eccentric anomaly & Mean of $E_b$ and $E_a$ \\
$E_H$ & Half eccentric anomalistic duration & One half of the difference between $E_b$ and $E_a$ \\
$\rho_*$ & Stellar density & Average density of the host star \\
$\zeta/R_*$ & Zeta over $R_*$ & The reciprocal of one half of the $T_{\mathrm{TS05}}$ duration. \\
$\Upsilon/R_*$ & Upsilon over $R_*$ & The reciprocal of one half of the $T_1$ duration \\
$\mathcal{D}$ & Deviation & Deviation of a candidate duration expression from the exact solution \\
$\mathcal{I}$ & Improvement & Improvement of a candidate duration expression over the $T_{\mathrm{TS05}}$ expression \\[1ex]
\hline\hline 
\end{tabular}
\label{table:nonlin} 
\end{table*}

\label{lastpage}


\begin{thebibliography}{99}
\bibitem[\protect\citeauthoryear{Bakos et al.}{2007}]{bak07} Bakos, G. \'{A}. et al., 2007, ApJ, 670, 826
\bibitem[\protect\citeauthoryear{Carter et al.}{2008}]{car08} Carter, J. A., Yee, J. C., Eastman, J., Gaudi, B. S. \& Winn, J. N., 2008, APJ, 689, 499
\bibitem[\protect\citeauthoryear{Einstein}{1915}]{ein15} Einstein, A., 1915, Preuss. Akad. Wiss. Berlin , 47, 831
\bibitem[\protect\citeauthoryear{Ford et al.}{2008}]{for08} Ford, E. B., Quinn, S. N. \& Veras, D., 2008, ApJ, 678, 1407
\bibitem[\protect\citeauthoryear{Fossey et al.}{2009}]{fos09} Fossey, S., Waldmann, I. \& Kipping, D., 2009, MNRAS, 396, 16
\bibitem[\protect\citeauthoryear{Hellier et al.}{2009}]{hel09} Hellier, C. et al., 2009, Nature, 460, 1098
\bibitem[\protect\citeauthoryear{Heyl \& Gladman}{2007}]{hey07} Heyl, J. S. \& Gladman, B. J., 2007, MNRAS, 377, 1511
\bibitem[\protect\citeauthoryear{Jord\'{a}n \& Bakos}{2008}]{jor08} Jord\'{a}n, A. \& Bakos, G. \'{A}. 2008, ApJ, 685, 543
\bibitem[\protect\citeauthoryear{Kipping}{2008}]{kip08} Kipping, D. M., 2008, MNRAS, 389, 1383 (K08)
\bibitem[\protect\citeauthoryear{Kipping}{2009a}]{kip09a} Kipping, D. M., 2009a, MNRAS, 392, 181
\bibitem[\protect\citeauthoryear{Kipping}{2009b}]{kip09b} Kipping, D. M., 2009b, MNRAS, 396, 1797
\bibitem[\protect\citeauthoryear{Kipping et al.}{2009c}]{kip09c} Kipping, D. M., Fossey, S. J. \& Campanella, G., 2009c, MNRAS, 400, 398
\bibitem[\protect\citeauthoryear{Kipping \& Tinetti}{2010}]{kip10} Kipping, D. M. \& Tinetti, G., 2010, submitted to ApJ
\bibitem[\protect\citeauthoryear{Kopal}{1959}]{kop59} Kopal, Z., 1959, Close binary systems (London: Chapman \& Hall)
\bibitem[\protect\citeauthoryear{Kozai}{1962}]{koz62} Kozai, Y., 1962, Astronomical Journal, 67, 591
\bibitem[\protect\citeauthoryear{Laughlin et al.}{2009}]{lau09} Laughlin, G., Deming, D., Langton, J., Kasen, D., Vogt, S., Butler, P. \& Rivera, E. 2009, Nature, 457, 562
\bibitem[\protect\citeauthoryear{Mandel \& Agol}{2002}]{man02} Mandel, K. \& Agol, E. 2002, ApJ, 580, L171
\bibitem[\protect\citeauthoryear{Miralda-Escud\'{e}}{2002}]{mir02} Miralda-Escud\'{e}, J., 2002, ApJ, 564, 1019
\bibitem[\protect\citeauthoryear{Murray \& Dermott}{1999}]{mur99} Murray, C. D., \& Dermott, S. F., 1999, Solar System Dynamics (Cambridge University Press)
\bibitem[\protect\citeauthoryear{Naef et al.}{2001}]{nae01} Naef, D. et al., 2001, A\&A, 375, 37
\bibitem[\protect\citeauthoryear{P\'{a}l}{2008}]{pal08} P\'{a}l, A. 2008, MNRAS, 390, 281
\bibitem[\protect\citeauthoryear{Seager \& Mall\'{e}n-Ornelas}{2003}]{sea03} Seager, S., \& Mall\'{e}n-Ornelas, G., 2003, ApJ, 585, 1038 (SMO03)
\bibitem[\protect\citeauthoryear{Seager \& Hui}{2002}]{sea02} Seager, S., \& Hui, L., 2002, ApJ, 572, 540
\bibitem[\protect\citeauthoryear{Sterne}{1939}]{ste39} Sterne, T. E., 1939, MNRAS, 99, 451
—. 1940, Proceedings of the National Academy of Science, 26, 36
\bibitem[\protect\citeauthoryear{Tingley \& Sackett}{2005}]{tin05} Tingley, B. \& Sackett, P. D., 2005, ApJ, 676, 1011 (TS05)
\bibitem[\protect\citeauthoryear{Winn et al.}{2009}]{win09} Winn, J. N. et al., 2009, ApJ, 703, 2091
\bibitem[\protect\citeauthoryear{Winn}{2010}]{win10} Winn, J. N., 2010, \emph{Transits and Occultations}, EXOPLANETS, University of Arizona Press; ed: S. Seager (W10)
\end{thebibliography}
\end{document}